\documentclass[a4paper]{article}
\usepackage{graphicx}
\usepackage{hyperref}
\newcommand{\etal}{{\it et~al.}}
\title{Electron-Hadron shower discrimination in a liquid argon time
  projection chamber} 
\author{J.J. Back,
G.J. Barker, 
A.J. Bennieston, 
S.B. Boyd, \\ 
B. Morgan and 
Y.A. Ramachers\footnote{e-mail: y.a.ramachers@warwick.ac.uk}\\
Dept. of Physics, University of Warwick, Coventry CV4 7AL, UK}
\date{}
\begin{document}
\maketitle
\begin{abstract}
By exploiting structural differences between electromagnetic and hadronic
showers in a multivariate analysis we present an efficient Electron-Hadron
discrimination algorithm for liquid argon time projection chambers, validated 
using Geant4 simulated data.
\end{abstract}

\section{Introduction}
\label{sec:Intro}
Liquid Argon Time Projection Chambers (LAr-TPCs) are recognised as being a
potential detector technology for a next-generation Neutrino oscillation
experiment and their development is the subject of vibrant R\&{}D programmes in
Europe, Japan and the USA (see Ref.~\cite{01} and references therein).
The simultaneous tracking and calorimetry capability, with millimetric
granularity, makes LAr-TPCs ideal to image Neutrino interactions over a wide
range of energies as demonstrated recently by ICARUS~\cite{icarus}. 

Despite the physics promise of LAr-TPCs, a fully automated (that is, with zero 
human interaction) software system for reconstructing events has proven 
difficult to achieve. With the availability of more and more event data,
recent years have 
seen progress in limited clustering of structures and tracking. However,
a general event is likely to contain a mixture of showers and tracks
originating from an unknown vertex anywhere in the sensitive volume. One major
challenge for automated reconstruction is to disentangle the  
fundamentally different shower and track structures and hence segment the 
event into its final state particles. Accurate segmentation is critical for 
subsequent particle identification on the substructures. Neutrino oscillation
analyses proposed for  
next-generation facilities rely heavily on accurate interaction
classification, motivating studies of this task.

In this article, we concentrate on the particle identification step, based on
extensive Monte-Carlo simulations, assuming 
exact segmentation, i.e. all hits belonging to each single particle have been
clustered with full efficiency and purity. We defer discussion of techniques
for this more challenging task to later publications.
We present a study using a set of effective and computationally simple 
variables which  
allow the discrimination of Electron shower structures from all relevant 
hadronic shower structures in a fine-grained tracking detector such as a 
LAr-TPC. For the intended physics application, the 
dominant Hadron shower structures result from Pions, 
Protons or Kaons in the final state. In this context, Electron-Hadron shower
discrimination is important for Neutrino oscillation studies in order to
measure  
Electron-neutrino events in a Muon-neutrino beam, resulting in an energetic 
Electron in the LAr medium after a charged-current interaction with 
a nucleus. Therefore, the Electron is considered to be the ``signal''
and anything non-electron, i.e. Hadron showers, is labelled as ``background'',
unless stated otherwise. 

Additionally, the energy resolution of any LAr-TPC will
depend on the nature of the particle depositing energy due to quenching
factors, i.e. particle dependent ionization efficiencies in liquid
argon. Electromagnetic and hadronic particle energy
depositions will each require separate energy calibrations depending on their
specific ionization as a function of energy, hence for their energy measurement 
particle identification would be beneficial.

The set of variables also prove to be useful for other particle identification 
tasks. Track structures associated with Protons or
Muons can be identified by these variables as well as, to a lesser extent, 
hadronic final states such as neutral and charged Pions and Kaons.

In the following, we describe the Monte-Carlo data production in order to
define our case studies, then the analysis algorithm by first introducing the
discrimination variables before finally presenting their applicability to
identifying particles in various Neutrino interaction event topologies.

\section{Monte-Carlo data production}
\label{sec:MC}
\begin{figure}[!htb]
\begin{center}
\includegraphics[width=0.8\textwidth]{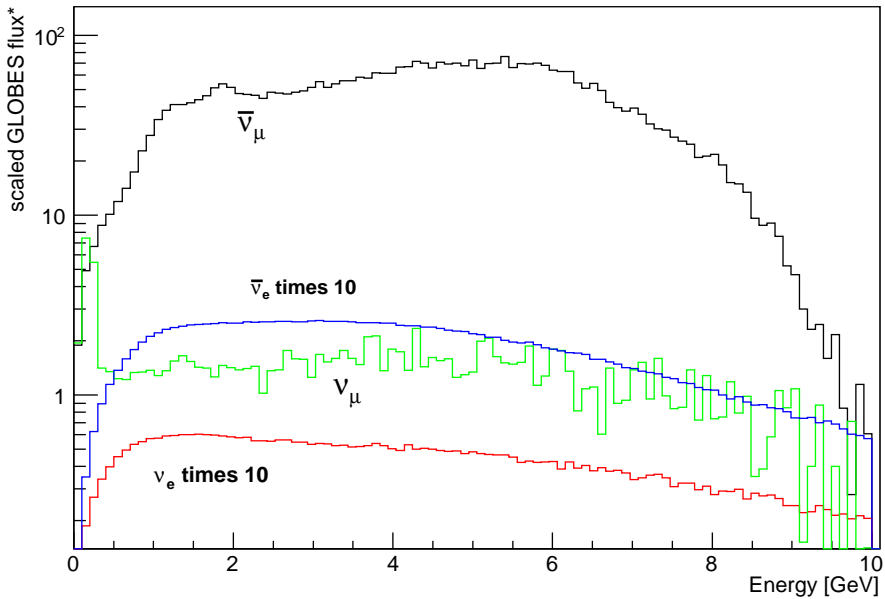}
\caption{The scaled CN2PY Neutrino beam fluxes~\cite{longhin} from
  negative Pion decays. Weighting factors for the fluxes are given in table
  \ref{tab_weighting}. *: Flux units are prepared for use in
  GLoBES ~\cite{globes} and scaled here by a factor
  $10^{-12}$. Parameters 
  entering the calculation of the correct GLoBES normalisation factor in order
to obtain SI flux units are: 3e21 p.o.t., a distance of 100 Km and an area of
100 m$^{2}$ with a detector at a distance of 2300 km. For details, see the GLoBES
user manual.} 
\label{fig_flux}
\end{center}
\end{figure}

Several simulated single particle event classes were prepared as input to the
analysis. We have tested the algorithm with a Muon--neutrino flux 
generated as part of the LAGUNA-LBNO FP7 programme~\cite{laguna} to evaluate
the physics potential of a long baseline experiment between CERN and an
underground 
far detector location in Pyhasalmi Finland (CN2PY). 
The CN2PY Neutrino flux~\cite{longhin} (un-oscillated) from decaying negatively 
charged Pions (for a distance of 2300\,km) provided the input for the first
case study. Neutrino reaction final states were produced using the GENIE event
generator (version 2.6.4)~\cite{genie}. The scaled Neutrino energy
distributions are shown 
in Fig.~\ref{fig_flux} with relative weighting factors given in
Table~\ref{tab_weighting}. 

\begin{table}[!htb]
\caption{The relative neutrino flavour weighting factors for the CN2PY
  Neutrino beam 
  (un-oscillated) from negatively charged Pions taken as input for this
 study. Note that the uniform Neutrino energy beam case study also takes these
flavour weighting factors into account.}
\label{tab_weighting}
\begin{center}
\begin{tabular}{cccc}
$\bar{\nu}_{\mu}$ flux & $\nu_{\mu}$ flux & $\bar{\nu}_{e}$ flux &
$\nu_{e}$ flux\\\hline
1.0 & $3.471\times{}10^{-2}$ & $4.455\times{}10^{-3}$& $1.0772\times{}10^{-3}$
\end{tabular}
\end{center}
\end{table}

The GENIE simulated events were un-filtered, with all
possible final state particles allowed. The total number of
each individual particle final state from GENIE in each of the four Neutrino
fluxes then results in individual weighting factors for that particle species
during the subsequent analysis. The complete set of weighting factors used
during the multivariate analyis, see below, is
listed in Tables~\ref{tab_particleW}, \ref{tab_particleW2}.  
 
These single particle energy distributions from the GENIE simulations then
serve as the 
particle source in the Geant4 (version 9.5.0) Monte-Carlo transport
simulation toolkit~\cite{geant4}. The simulation models an
homogeneous liquid argon detector as a cylinder with a radius and height of
100\,m, in order to fully contain the showers. Particles start at the centre
of the cylindrical volume and are tracked through the volume with
interactions modelled by the QGSP\_BIC\_HP physics list, with all
electromagnetic 
and hadronic processes enabled, as recommended by the Geant4
collaboration for our energy range of interest. The LAr volume is divided into
$(1\times{}1\times{}1)$ mm$^{3}$ cubic voxels, and all primary and secondary
particles are tracked through voxels down to zero energy, or until they leave
the volume. Energy deposits by charged particles passing through voxels are
tallied into a map between voxel coordinates and the total energy
deposited. No attempt was made to model the readout system, but 
quenching of ionisation events specific to LAr detectors was taken into
account following \cite{quench}. 

Results are collected and processed in a
final step by four selected multivariate analysis algorithms (MVA) taken
from the TMVA toolkit~\cite{tmva}. Each MVA is based on supervised learning,
requiring separate training and 
test data samples. Each sample is chosen from a random
split in half of the available data. The final evaluation delivers
optimal cuts for each selected MVA method, producing efficiencies for
signal selection, background selection and corresponding purities. Note that
the complete set of weighting factors as
listed in Tables~\ref{tab_particleW}, \ref{tab_particleW2} is used in each MVA
method. The chosen methods are: the k--nearest neighbour search (KNN), two
variations of boosted decision trees (BDT, BDTG) and a neural network
(MLPBNN). These four represent the most successful out of a substantially
larger number of methods in the TMVA toolkit during preliminary tests on our
project. 

Our second case study attempts a pure algorithm test by running the identical
analysis as discussed above over a sample of final state particles sampled 
from a uniform Neutrino energy beam (same flavour weighting factors, same
energy range as the CN2PY case). This case studies the impact of a less well
constrained neutrino beam energy distribution.

\begin{table*}[!htb]
\caption{The relative weighting factors for the final state
  particles resulting from the CN2PY Neutrino beam simulation containing
  particles given 
  in the first column. These weighting factors are each taken into
  account during the multivariate analysis stage and calculated by multiplying
  relative simulated final state particle numbers with the corresponding
  neutrino 
  flavour weighting factors (see the discrepancy on magnitude between
  $\bar{\nu}_{\mu}$ values and
  all other weighting factors due to the small neutrino flavour weighting
  factors). Note that Muon and
  Electron energy spectra are taken from Genie simulations, also using the
  corresponding neutrino flavour weighting factor.}
\label{tab_particleW}
\begin{center}
\begin{tabular}{l|ccccccc}
 & $\pi^{+}$ & $\pi^{-}$ & $\pi^{0}$ & p & $K^{\pm}$ & $K^{0}$& $\gamma$\\\hline
$\bar{\nu}_{\mu}$ & 0.203& 0.508& 0.409& 1.610& 0.016& 0.022& 0.035\\
$\nu_{\mu}$       & $2.42 \times 10^{-2}$& $9.89 \times 10^{-3}$& $2.04 \times 10^{-2}$& $6.284 \times 10^{-1}$& $1.43 \times 10^{-3}$& $1.1 \times 10^{-3}$& $1.96 \times 10^{-3}$\\
$\bar{\nu}_{e}$  & $9.4 \times 10^{-4}$& $2.32 \times 10^{-3}$& $1.85 \times 10^{-3}$& $7.2 \times 10^{-3}$& $7.7 \times 10^{-5}$& $1.1 \times 10^{-4}$& $1.8 \times 10^{-4}$\\
$\nu_{e}$        & $7.6 \times 10^{-4}$& $3.1 \times 10^{-4}$& $6.4 \times 10^{-4}$& $1.95 \times 10^{-3}$& $4.5 \times 10^{-5}$& $3.4 \times 10^{-5}$& $5.7 \times 10^{-5}$\\
\end{tabular} 
\end{center}
\end{table*}

\begin{table*}[!htb]
\caption{Same as table \protect{\ref{tab_particleW}} but for the second case study
  of a uniform neutrino energy beam.}
\label{tab_particleW2}
\begin{center}
\begin{tabular}{l|ccccccc}
 & $\pi^{+}$ & $\pi^{-}$ & $\pi^{0}$ & p & $K^{\pm}$ & $K^{0}$& $\gamma$\\\hline
$\bar{\nu}_{\mu}$ & 0.251& 0.567& 0.466& 1.635& 0.023& 0.029& 0.046\\
$\nu_{\mu}$       & $2.61 \times 10^{-2}$& $1.13 \times 10^{-2}$& $2.25 \times 10^{-2}$& $6.32 \times 10^{-2}$& $1.68 \times 10^{-3}$& $1.28 \times 10^{-3}$& $2.13 \times 10^{-3}$\\
$\bar{\nu}_{e}$  & $1.1 \times 10^{-3}$& $2.54 \times 10^{-3}$& $2.1 \times 10^{-3}$& $7.29 \times 10^{-3}$& $9.9 \times 10^{-5}$& $1.4 \times 10^{-4}$& $1.9 \times 10^{-4}$\\
$\nu_{e}$        & $8.1 \times 10^{-4}$& $3.5 \times 10^{-4}$& $7.1 \times 10^{-4}$& $1.97 \times 10^{-3}$& $5.4 \times 10^{-5}$& $3.9 \times 10^{-5}$& $6.7 \times 10^{-5}$\\
\end{tabular}
\end{center}
\end{table*}
\section{Electron-shower discrimination algorithm}
\label{sec:ElecShower}
Below, we detail the calculation and physics motivation of four discriminating 
variables. The main assumption underpinning all of the following is that
clustered energy deposits, hits, in the detector from a particular particle
have been collected as a single object by the reconstruction, i.e. a
cluster. Results are collected and processed in a 
final step by four selected multivariate analysis algorithms (MVA) taken
from the TMVA toolkit~\cite{tmva}. This final step is discussed in more detail
in section \ref{sec:MC}. 

\begin{figure*}[!htb]
\begin{center}
\includegraphics[width=0.8\textwidth]{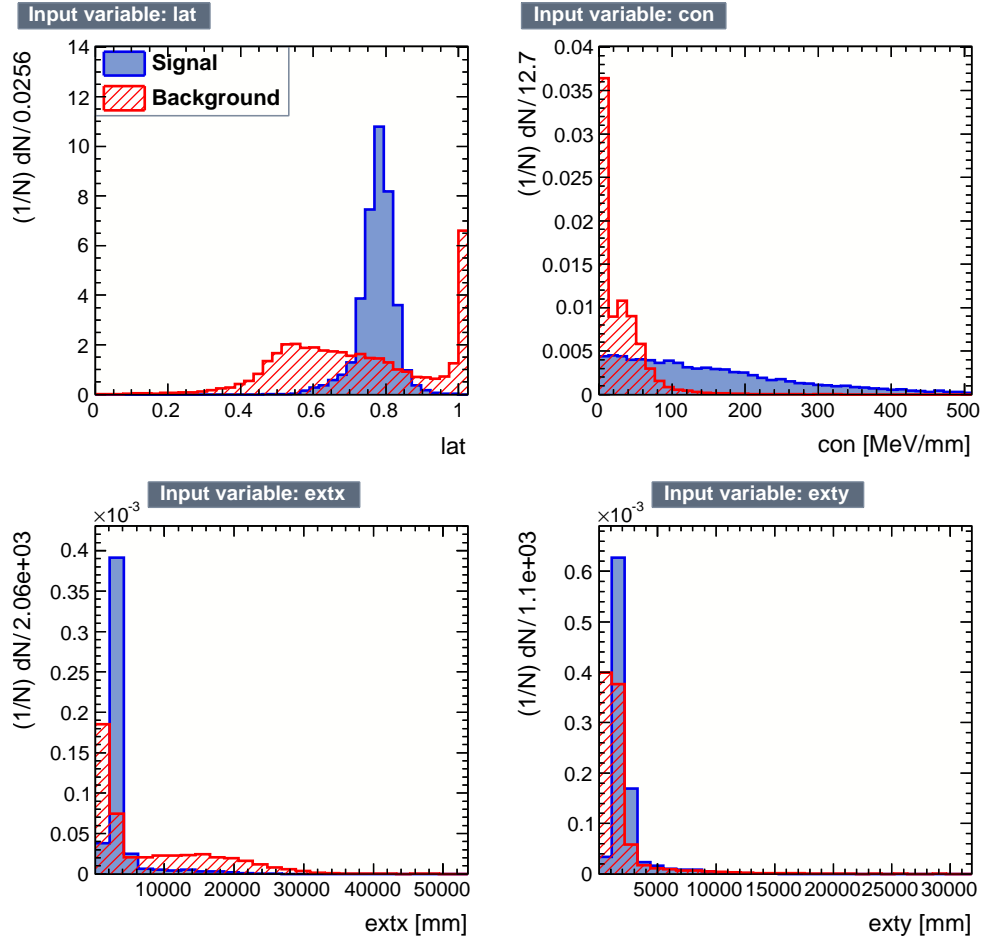}
\caption{The set of four discrimination variables applied to the CN2PY beam
  (see text) 
  Neutrino events. Signal events are defined as electromagnetic shower events
  from Electrons, compared to the complete shower background, which 
  includes all simulated structures from Protons, charged Pions, neutral Pions,
  charged and neutral Kaons and Muons, all with appropriate weighting
  factors. For MVA training purposes the event sample size is set to
  $10^{4}$ for each individual signal and background contribution. The
  histograms for signal and background are scaled individually to equal
  integrals for display purposes only.}
\label{fig_hev}
\end{center}
\end{figure*}
\begin{figure*}[!htb]
\begin{center}
\includegraphics[width=0.8\textwidth]{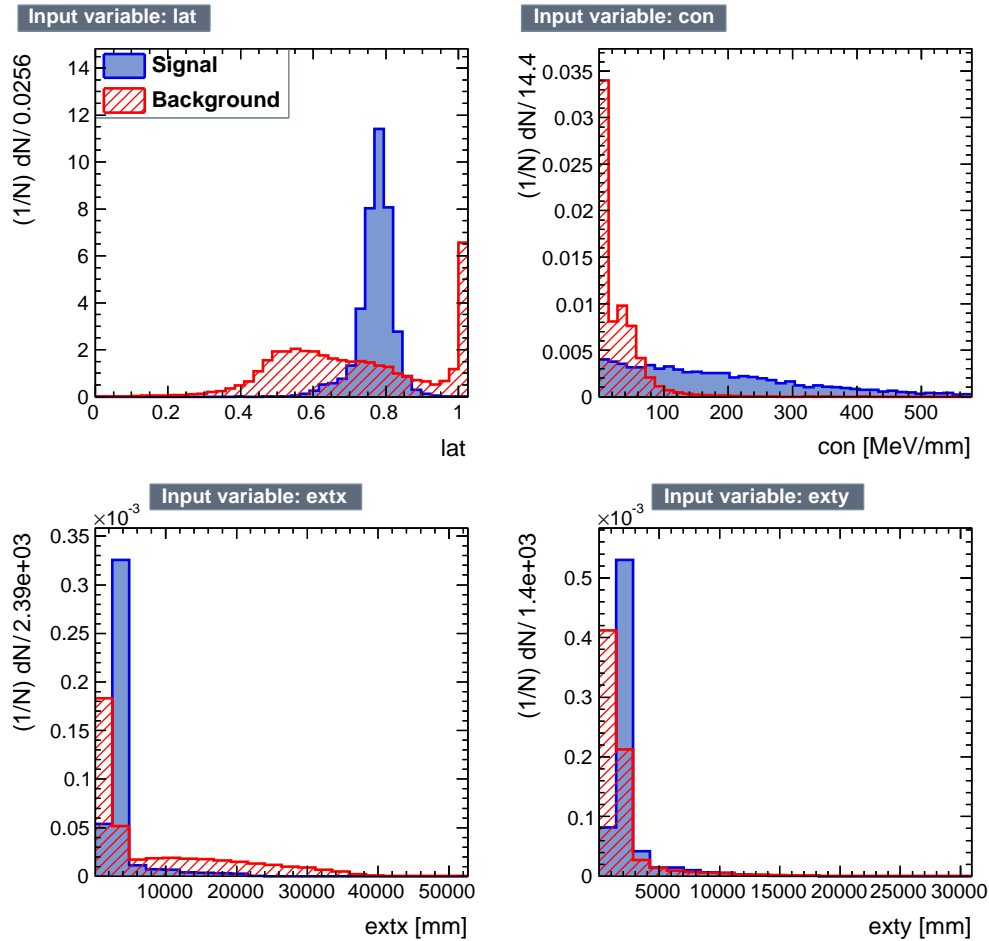}
\caption{Similar to Fig.~2; the set of four discrimination variables 
  but for the second case study of a uniform neutrino energy beam.} 
\label{fig_lev}
\end{center}
\end{figure*}
This crucial clustering step for any LAr detector analysis
is assumed to have taken place already. It should be stressed that a
reliable, complete and automatic initial clustering step is, in fact, among the
unsolved challenges of LAr data analysis, owing to the richness of
topological event structures in such a detector, and will not be addressed here.

The first step of the discrimination algorithm consists of transforming the
isolated cluster using a principal component analysis (PCA)~\cite{pca}. The data
at this stage consists purely of detector hits forming a
point-cloud of unknown shape. Dealing with LAr detectors translates
into the capability of assigning three spatial coordinates and a charge value to
each individual hit (with corresponding uncertainties). 

\begin{figure*}[!htb]
\begin{center}
\includegraphics[width=0.5\textwidth]{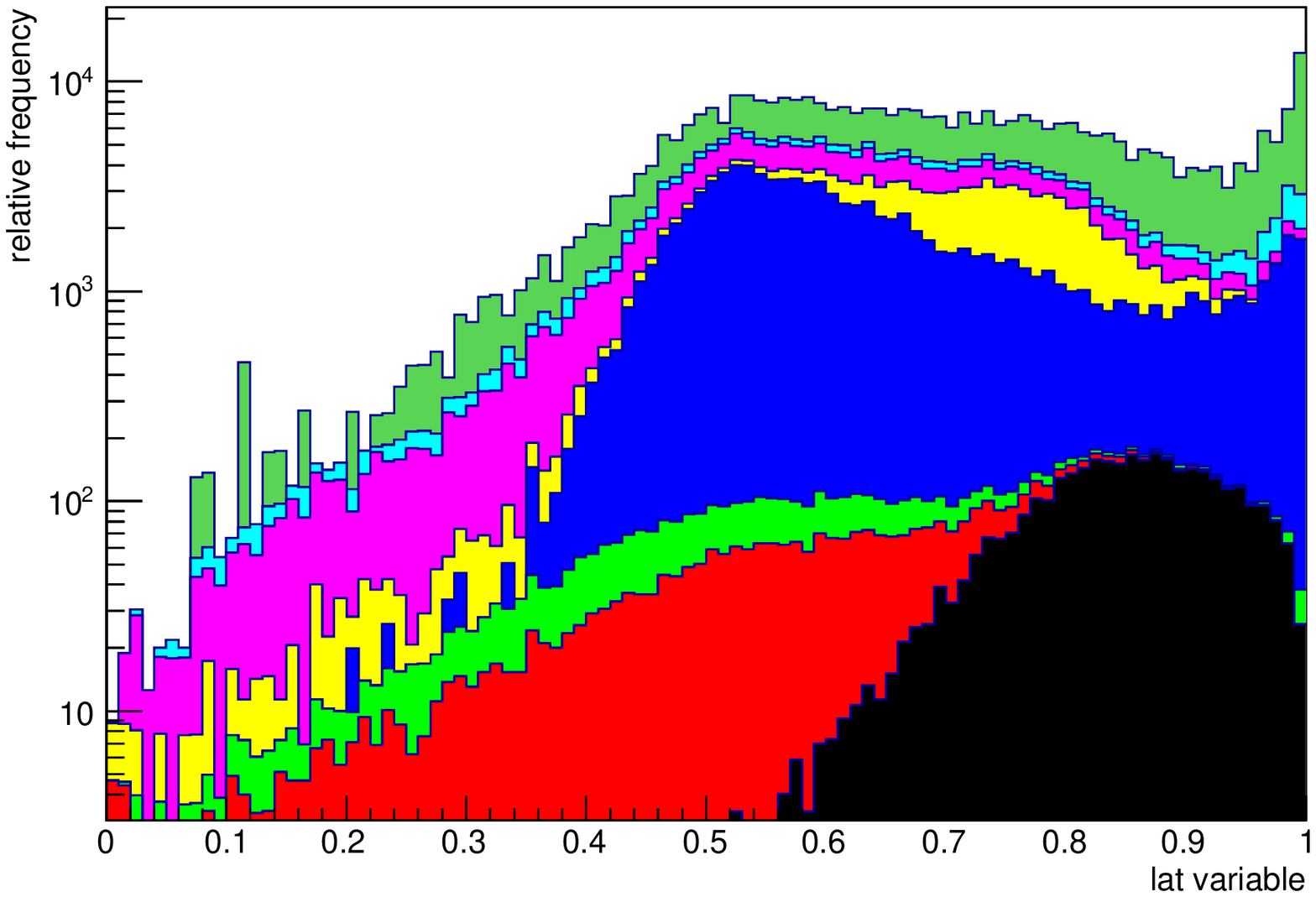}\hspace*{6pt}
\includegraphics[width=0.5\textwidth]{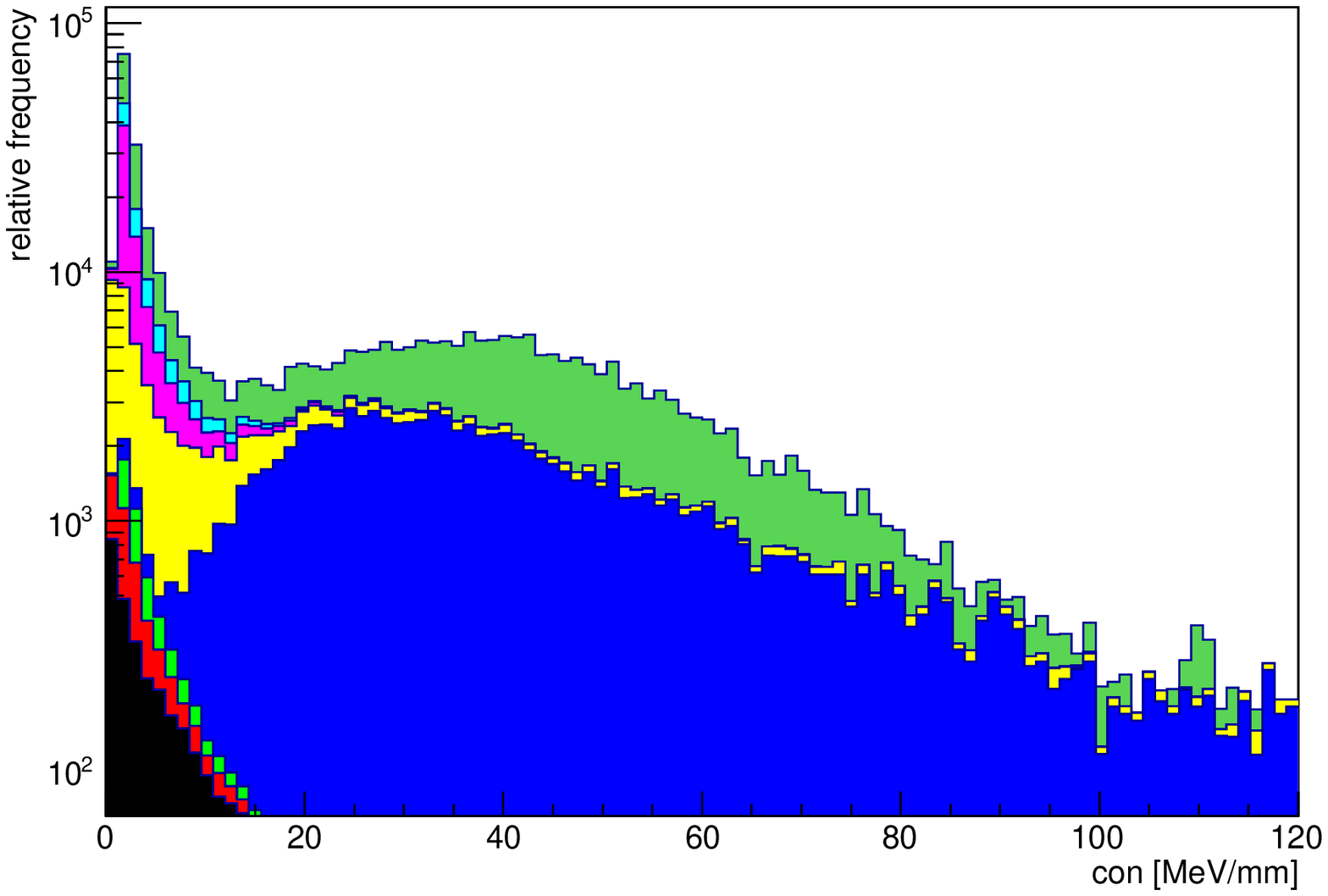}
\caption{Stacked histogram example displaying the background composition due
  to simulated final state particles from the CN2PY neutrino beam case. This
  example shows the background in the 'lat' variable (left panel) for an
  electron 
  signal for the dominant anti-muon neutrino flavour and the 'con' variable on
  the right panel. The individual
  histograms are normalised and scaled subsequently by their corresponding
  weighting factor (particle dependent) from table~\ref{tab_particleW}. The
  additional neutrino flavours contributing to that background in Fig.~1 are
  suppressed to avoid clutter from too many colours on the plot. The colour
  coding is as follows: black - gamma, red - K$^{0}$, green - charged Kaons,
  blue 
  - Muon, yellow - neutral Pion, pink - Pi$^{-}$, light blue - Pi$^{+}$, dark
  green - Proton.} 
\label{fig_lev2}
\end{center}
\end{figure*}

The reason for the initial PCA step is that it allows us to consider the unknown
point-cloud to be aligned and ordered in a consistent way. The PCA
is set up such that it calculates the three principal spatial components
(ignoring 
the charge of the hit) as eigenvalues of the transformation matrix, which
itself consists of the corresponding eigenvectors. Applying the transformation
aligns the coordinate axes to the principal axes, using the convention that the
dominant principle axis becomes the 'x'-axis in a right-handed Cartesian
coordinate 
system. This allows to utilise
transverse and longitudinal geometric information due to the presence of a well
defined axis for the structure.

\subsection{The lateral projection variable ``lat''}
\label{sec:Lat}
This discrimination variable is motivated by previous and ongoing accelerator
experiments and their analysis of electromagnetic (EM) and Pion showers (see for
example Ref.~\cite{h1}). The entire point-cloud is projected onto the plane of
minor principal components (the lateral plane to the major principal
axis). A core region is defined around the projected mean value and the ratio
of energy (charge) deposited in the core to the total energy (charge) yields
the ``lat'' variable. 

Charged Hadron showers feature either a dense core region, often resembling a
track-like structure with only a thin ``halo'' around it, or  
dissolve into a wide-spread, loose cloud of hits. Proton tracks might or might
not shower 
at all. EM showers, by contrast, tend to produce more constant halo-to-core
ratios, as shown in Fig.~\ref{fig_hev}, \ref{fig_lev} and \ref{fig_lev2}.

The precise definition of the core region turns out to be non-critical. For
simplicity, 
a circular shape is chosen, which is centred on the mean value of the 
projected point-cloud and is defined
with half the Moli\`{e}re radius in liquid argon, which is equal to 
9.61\,cm~\cite{aprile}. Each hit inside the core contributes to the sum of
energy 
in the core, which is then divided by the total energy of the shower (the sum of
the energies of all hits).

\subsection{The hit concentration variable ``con''}
\label{sec:Con}
While the previous variable emphasises a subdivision of the shower structure
into two main components (core and halo), the hit concentration variable ``con''
attempts to use the same lateral information globally. This is independent of
any material constant, such as the Moli\`{e}re radius.

The variable first calculates the radial distance of each hit from the main
principal axis, i.e. the radial distance from the origin in the lateral
projection plane. A minimum threshold radius is set in order to avoid the
Coulomb singularity (for practical purposes chosen as 1\,mm in this
analysis), and finally the hit concentration is calculated as the sum of charge
(energy) of the hits divided by their radial distance.

The idea is to calculate the
Coulomb potential energy of the point-cloud around the main principal
axis. This approach delivers complementary information to the previous rigid
division into two parts, since it emphasises the extremes such as track-like
structures as being particularly highly concentrated, and very loose structures
as being particularly low in concentration. Figures~\ref{fig_hev},
\ref{fig_lev} and \ref{fig_lev2} 
show the main benefit of this variable. The Hadron showers are dominated by
less densely concentrated structures (they peak 
at low Coulomb energies around the main principal axis). A 
second population of Hadron events contain
Proton tracks as well as sufficiently dense Pion tracks, contributing a larger 
value to this concentration variable. EM showers are rather non-specific but
are significantly more concentrated than Hadron showers. 

\subsection{Spatial extent variables ``extx,y''}
\label{sec:Ext}
The final two proposed discrimination variables are obtained by quantifying
the spatial extent of the hit cloud in the three available principal component
axes. A definition of the extent in purely geometrical terms turns out to be
very useful. The algorithm calculates extent in all three available axes but
one of the two transverse axes is discarded since it does not contribute any
new information. This azimuthal symmetry assumption around the dominant
principal axis was extensively tested and confirmed during this study.

The spatial extent is calculated by obtaining the convex hull in 3D around the
hit cloud and finding the maximum distance (extent) of the hull on all
three axes. Building the convex hull around an object is a standard algorithm
in image analysis \cite{hull}. Even though its application in 3D is rather
exotic, 
ready-made implementations of the code exist, for instance in the scientific
python library, SciPy~\cite{scipy}. The meaning of the term convex hull can be
visualised by considering stiff packaging paper, wrapped around an irregularly
shaped object, i.e. the hit cloud, in order to enclose it completely. The
paper would represent the convex hull, without the folds and overlaps
real paper would produce.

The convex hull is obtained by calculating the Delaunay tessellation
\cite{hull} of the 
hit cloud and finding the outermost Delaunay triangles, made of data members, 
enveloping the entire cloud. The difference between the maximum and minimum 
coordinate value on each (principal component) coordinate axis is calculated,
giving the extent of the hull in all three dimensions.

\section{Results}
\label{sec:Results}

\begin{figure*}[!htb]
\begin{center}
\includegraphics[width=0.5\textwidth]{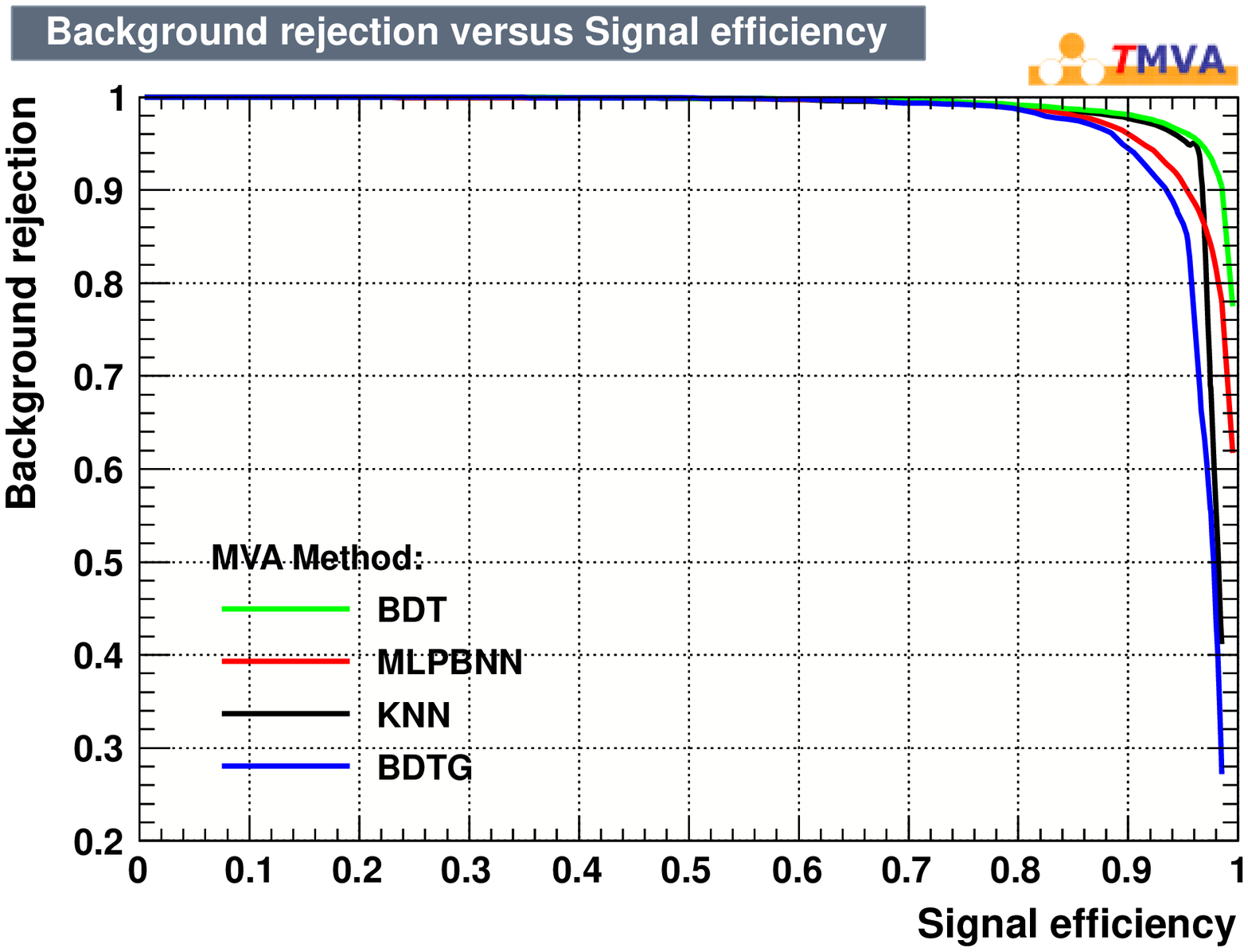}\hspace*{6pt}
\includegraphics[width=0.5\textwidth]{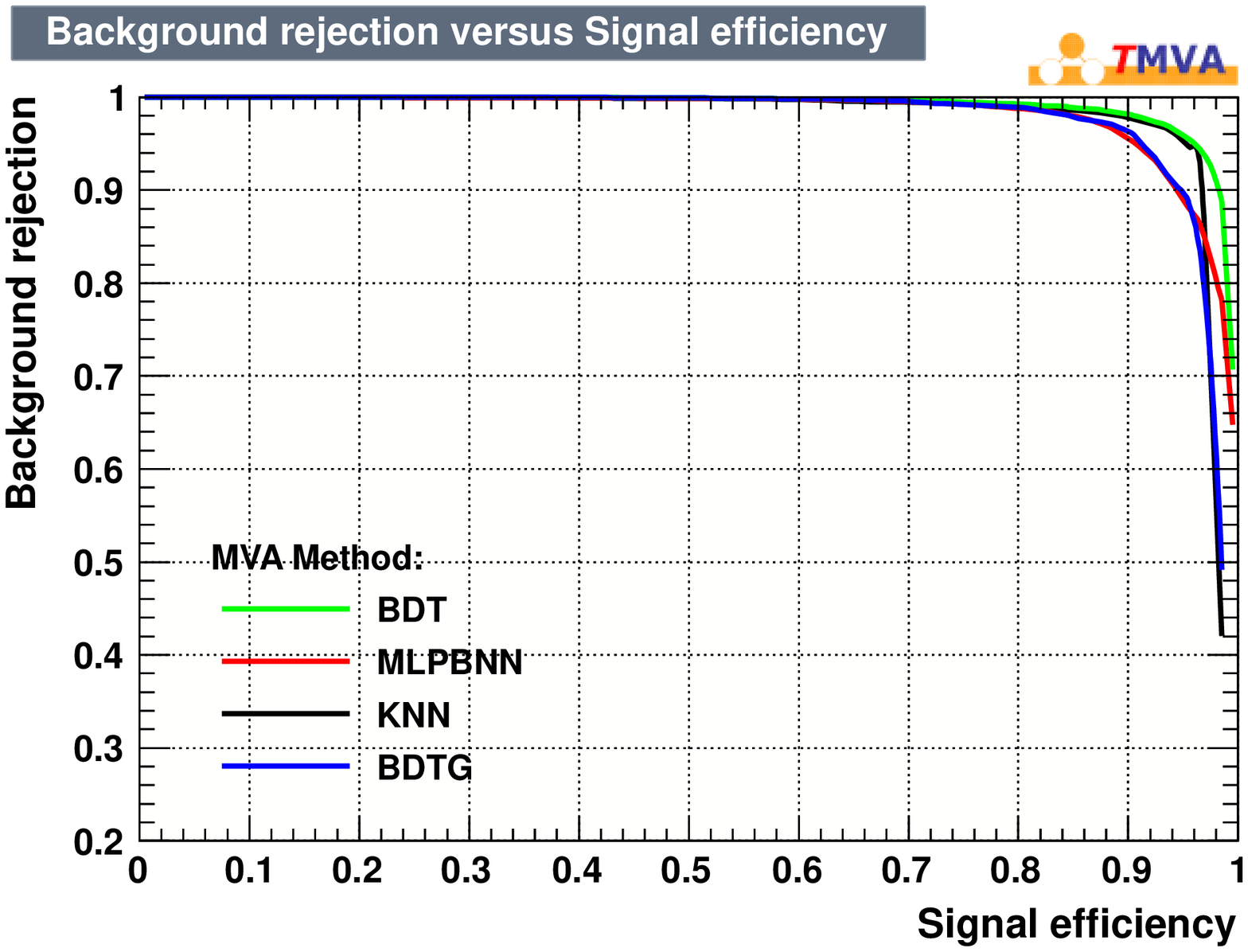}
\caption{Background rejection as a function of the signal
efficiency for all four chosen methods for an EM shower signal in a (left
panel) CN2PY beam and 
(right panel) uniform neutrino energy distribution. The methods are: the
k--nearest neighbour search (KNN), two 
variations of boosted decision trees (BDT, BDTG) and a neural network
(MLPBNN), all originating from the TMVA toolkit~\cite{tmva}. 
For numerical results on this analysis, see Tables~\ref{tabEMfull},
\ref{tabAllfull}.}
\label{fig1}
\end{center}
\end{figure*}

\subsection{Electron to combined Hadron/lepton background discrimination}
\label{sec:elecResults}
Results in this section summarise the performance of the proposed discriminating
variables when attempting to identify Electrons in a data sample consisting of
all dominant final state particles, i.e. charged and neutral Pion, Protons,
Kaons and Muons. 
The most straightforward way of illustrating the effectiveness of the proposed
variables can be seen in Fig.~\ref{fig1}, showing the background rejection as
a function of signal
efficiency for all four chosen methods, the nearest neighbour (KNN), two
variations of boosted decision trees (BDT, BDTG) and a neural network
(MLPBNN). These plots would ideally show a complete background rejection for all
signal efficiencies up to 100\%. The closer the curves follow these ideal
lines on the figure, the better the discrimination of the selected MVA
method. The MVA methods mentioned above have been chosen as the most effective
among a list of 24 possible methods offered in the 
TMVA toolkit. 

\begin{figure*}[!htb]
\begin{center}
\includegraphics[width=0.8\textwidth]{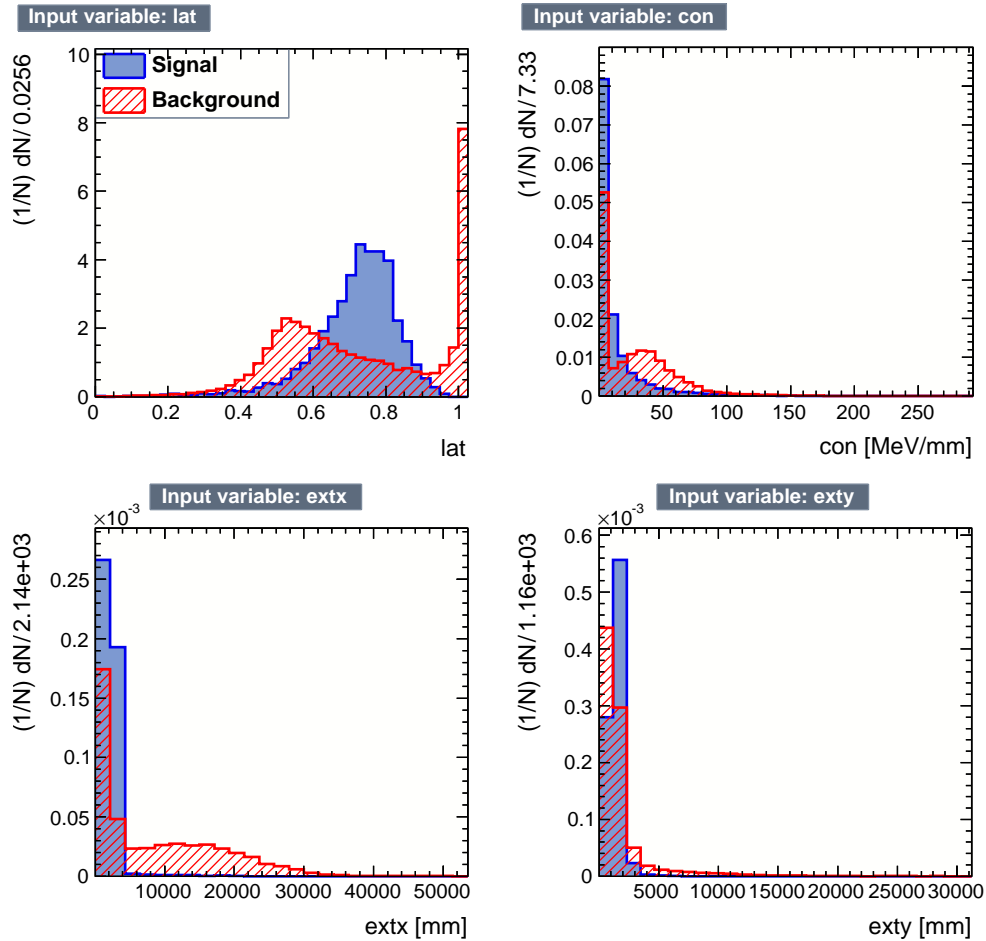}
\caption{Similar to Fig.~2 for the CN2PY beam producing a $\pi^{0}$ final
  state particle signal compared to the complete shower background.} 
\label{fig_hevpi}
\end{center}
\end{figure*}

Quantitative results for efficiencies and purities in EM shower discrimination
are given in Table~\ref{tabEMfull}, and illustrate a successful
(well above 90\% signal efficiency and background rejection) analysis
technique in 
distinguishing EM showers from Hadron showers in a LAr detector at
all interesting Neutrino energies. Finally, the method appears
relatively insensitive to presence or absence of a priori beam knowledge.

\subsection{Additional applications}
\label{sec:applic}
The variables discussed above lend themselves to further applications simply
because they are sensitive to only the geometrical shapes of clouds of 
hits in 3D space as opposed to kinematic variables. As a first application we
consider neutral current (NC) interactions in LAr. NC interactions form an
important background to Neutrino oscillation signals since they can produce
neutral pions in an event. Due to their prompt decay into a
pair of photons, with the inevitable EM showers, they constitute a 
problematic background to Electron-neutrino charged-current interactions. This
means that identifying final state 
$\pi^{0}$ would be beneficial for several reconstruction tasks. One study
\cite{icaruspi0} delivered a strong discrimination factor of $10^{-3}$ for
mis-identified $\pi^{0}$ in a LArTPC, a number which is widely quoted. Our
efforts here deliver a weaker $\pi^{0}$ discrimination, see below. However it
does not rely on complex computational tasks such as shower separation and
fitting and subsequent vertex finding in a potentially very busy event
structure, i.e. the reason why \cite{icaruspi0} relies on visual event
selection. 

The optimal signal and background efficiencies for the $\pi^{0}$ selection 
are shown in Table~\ref{tabAllfull} (see also Fig.~\ref{fig_hevpi}). Just as
for the study on Electron to Hadron
discrimination, all subsequent studies on changing the signal, here to the
$\pi^{0}$ particle, consider the complete set of background particles and
weighting factors, listed in Table~\ref{tab_particleW} and additionally the
relevant leptons, i.e. Muons and Electrons. All final state
particles other than the selected signal particle constitute background.

Other final state particles can also be identified as signals with varying
degrees of 
success. In Table~\ref{tabAllfull} we list results for Muons, Protons,
Kaons and $\pi^{+}$, where $\pi^{-}$ results are practically identical to Kaon
results, hence have been left out.
\begin{table*}[!htb]
\caption{Particle identification efficiencies and contamination (background
  efficiency or 'contamination' is equal to 1.0 - background rejection) on
  electron shower detection in a combined
  background of Hadron showers and Muons for all selected MVA
  methods. Weighting factors for each final state particle are always taken
  into account for signal as well as background particles. Efficiencies are
  calculated for the optimal decision boundary
  values defined by maximising the signal-to-total event ratio,
  $S/\sqrt{(S+B)}$.}
\label{tabEMfull}
\begin{center}
\begin{tabular}{ll|c|c}
\hline
MVA & Beam  & Electron--Signal/Background efficiency [\%] & Signal-to-total
$S/\sqrt{(S+B)}$\\
method & case study & & \\
\hline
\hline
KNN & CN2PY & 95.7 / 5.2 & 30.1\\
MLPBNN & CN2PY & 93.9 / 7.4 & 29.5\\
BDT & CN2PY & 96.2 / 4.5 & 30.3\\
BDTG & CN2PY & 90.6 / 6.0 & 29.2\\\hline
KNN & Uniform $\nu$ & 95.4 / 5.0 & 30.1\\
MLPBNN & Uniform $\nu$ & 92.9 / 7.3 & 29.4\\
BDT & Uniform $\nu$ & 96.1 / 4.9 & 30.2\\
BDTG & Uniform $\nu$ & 90.6 / 4.2 & 29.4\\
\end{tabular}
\end{center}
\end{table*}

\begin{table*}[!htb]
\caption{Particle identification efficiencies and contamination (background
  efficiency or 'contamination' is equal to 1.0 - background rejection) in a
  combined 
  background of Hadron showers and Leptons for the best MVA
  method, boosted decision trees, BDT. Efficiencies are calculated for the
  optimal decision boundary 
  values defined by maximising the signal-to-total event ratio,
  $S/\sqrt{(S+B)}$. Signal particles are listed in the top row.}
\label{tabAllfull}
\begin{center}
\begin{tabular}{ll|c|c|c|c|c}
\hline
MVA & Beam  & \multicolumn{5}{c}{Particle Signal/Background efficiency [\%]}\\
method & case study & $\pi^{0}$& Muon& Proton& Kaon & $\pi^{+}$\\
\hline
BDT & CN2PY & 96.8 / 15.9& 97.6 / 1.4& 93.7 / 28.3& 93.4 / 24.0& 97.3 / 41.7\\
BDT & Uniform $\nu$ & 96.1 / 15.6 & 98.1 / 2.3 & 92.8 / 29.9 & 92.2 / 25.1 & 97.9 / 43.1\\
\end{tabular}
\end{center}
\end{table*}

\begin{figure*}[!htb]
\begin{center}
\includegraphics[width=0.8\textwidth]{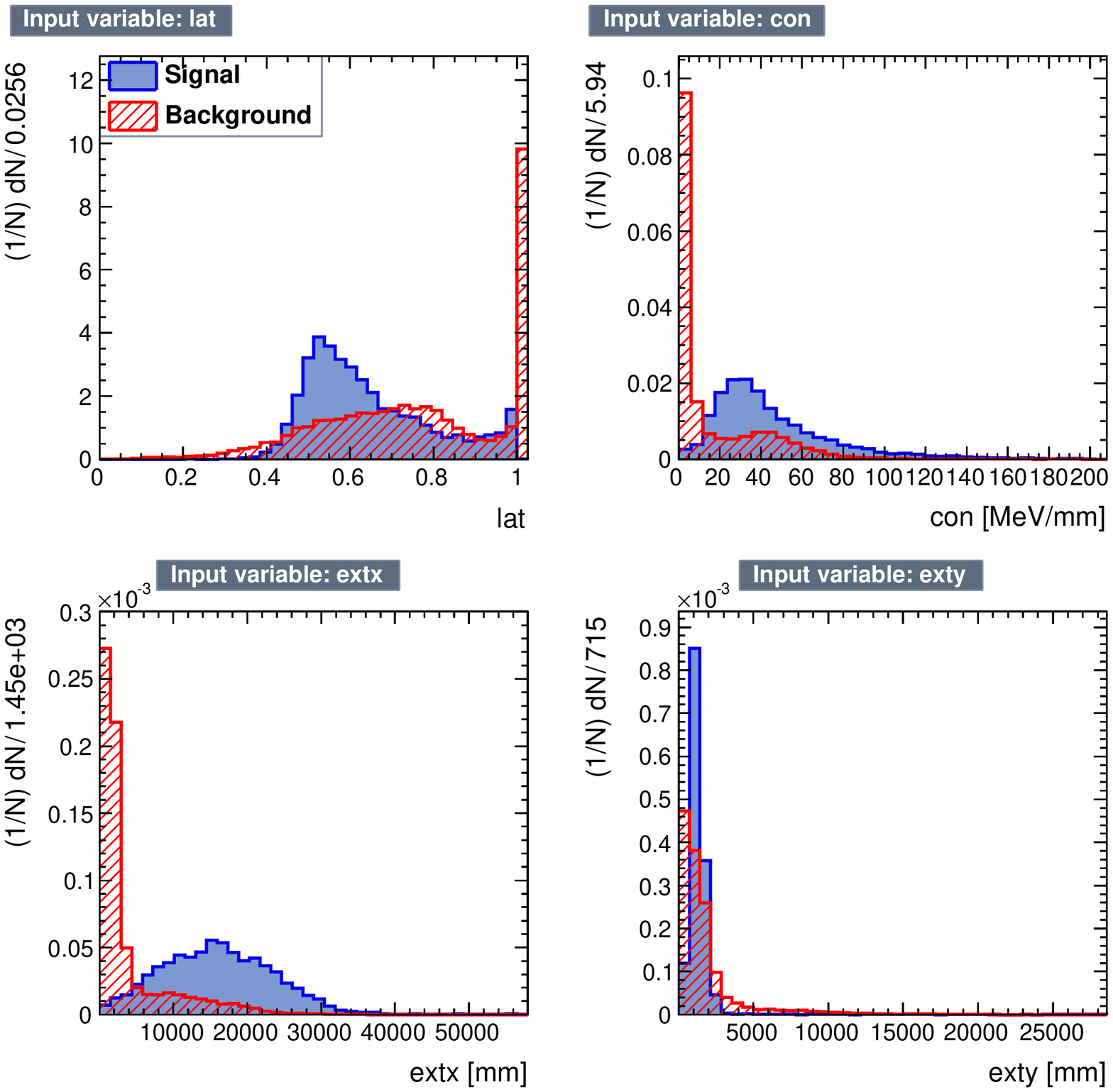}
\caption{Similar to Fig.~2 for the CN2PY beam producing Muons in the final
state, here defined as signal compared to the complete shower background.} 
\label{fig_mu}
\end{center}
\end{figure*}
\subsection{Detector threshold effects}
\label{sec:Background}
Detector-specific effects depend heavily on our primary
assumption of successful clustering of hits belonging
to a single particle in an overall unspecified detector event. However, any
automated analysis will at some point have to solve this challenge for
finely-grained LAr detector devices. 

Nevertheless, threshold variations would remove (or add) hits to the total
simulated single particle event and potentially change geometrical
structures. In order to test the robustness of our method against such a
realistic detector effect, we studied two scenarios: one realistic hit energy
cut at 120\,keV/mm for 
quenched (measured) energies and one essentially without an hit energy cut 
(1\,keV/mm). The threshold choice of 120\,keV/mm originates from the simulated
onset of the Muon energy loss peak which would be left practically fully
intact by this cut. The minimal cut scenario at 1\,keV/mm roughly doubles the
total amount of hits for most single particles simulated. 

The change from these threshold effects on 
all of the particle identification results shown earlier is negligible, 
typically less than 0.3\% on any signal or background efficiency quantity.

Similar robustness should be pointed out with respect to hit smearing effects,
i.e. the case when individual hits are represented by more realistic volumes
of three to six mm cubes (or even asymmetric dimensions). Our proposed
variables are purely measures of the geometry of structures in a LArTPC. As
long as the smallest spatial measure, i.e. a hit, is small compared with the
extent of shower/track structures, we do not expect any of our analysis
results to change significantly. 

\section{Conclusion}
\label{sec:Conclusion}
We have presented a study identifying particles in event structures obtained in
high-granularity detectors such as a liquid argon TPC using four 
discriminating variables. Combining them using MVA algorithms from the
TMVA toolkit gives signal efficiencies for Electron and Muon identification
well above 90\% with high background rejection efficiencies 
on all Neutrino interaction channels. Protons, Kaons and charged Pions show
anything from
acceptable to poor background rejection but high signal efficiency. Neutral
Pion identification can 
be made very efficient if a background contamination  of around 16\% is not a
concern. These 
results depend on the two
specific case studies undertaken here: first for a possible neutrino beam from
CERN to Finland and second, for a similar neutrino beam (same flavour
weighting factors) but uniform neutrino energy distribution. 

The data used in this work has been
assumed to consist of hits containing three spatial coordinate
values and one energy value, which should be the case for LAr TPC
detectors. 
Furthermore, it is assumed that each
single structure to be analysed has been clustered correctly which is still an
outstanding 
challenge for automated data analysis on high granularity detector data.
Nevertheless, if such a clustering can be achieved by some means, then
this work proposes an efficient and reliable method to identify Electron (as
well as Positron) 
events in a combined hadronic and leptonic background. These would be the
primary signatures of charged current Electron--Neutrino interactions which
constitute the main target of future long baseline experiments.
It also allows the identification of neutral Pions without additional
knowledge of decay vertices 
(or similar quantities) in a mixed background of other Hadrons and Leptons, and
can be used for Muon identification and as an efficient first pass on Proton,
Kaon and charged Pion identification. 

\section*{Acknowledgements}
We acknowledge the contribution made to this study by Timothy
Hughes and Laurence Woodward who worked with the corresponding author as final
year project students on this topic. Support on the STFC consolidated grant
ST/H00369X/1 is also acknowledged.

\end{document}